# Mitigating Spreadsheet Model Risk with Python Open Source Infrastructure

*Oliver Beavers*
*Director, Trivium Financial Group*
oliver@triviumfinancialgroup.com
*969 2nd St. SE, Charlottesville, VA 22902*

**ABSTRACT**

*Across an aggregation of EuSpRIG presentation papers, two maxims hold true: spreadsheets models are akin to software, yet spreadsheet developers are not software engineers. As such, the lack of traditional software engineering tools and protocols invites a higher rate of error in the end result. This paper lays ground work for spreadsheet modelling professionals to develop reproducible audit tools using freely available, open source packages built with the Python programming language, enabling stakeholders to develop clearly defined model "oracles" with which to test and audit spreadsheet calculations against.*

## 1. INTRODUCTION

Increasingly, publications focused on spreadsheet risk – whether from EuSpRIG, or project finance modelling books – have begun to lean towards a modelling methodology that uses combinations of array formulas and array-level range names [Swan, 2016]. Other approaches to spreadsheet modelling such as the FAST standard propose otherwise, favoring an arguably more readable approach with the use of names only as they apply to external links[FAST, 2017].

The human error of spreadsheet modelling has been well established. EuSpRIG's Horror Stories [EuSpRIG, 2017] provide accounts of significant financial loss due to spreadsheet error. Additional research has further recognized these errors, and makes calls for additional testing, particularly in larger spreadsheet models where the likelihood of at least one per-cell error is significantly higher [Panko, 2015].

Outside of the spreadsheet community, the Python programming language has grown significantly in user adoption in large part due to its easy to use syntax, and the strength of the open source computing libraries built on top of it. One of the more recent innovations for the language is that of the Jupyter Notebook: a highly interactive computing environment with a user-friendliness similar to that of Excel's.

Though Python's modelling applications have traditionally focused on quant finance and data science, the user-friendliness of its Jupyter Notebook interface, high quality computing and code-testing packages, and (as of recent) strong integration with Excel, make it a prime candidate for replicating modules of detailed financial models, which may be developed once, and reused in a fashion

This paper defines an introductory approach for spreadsheet professionals to learn to utilize Python's numerical capabilities by creating a simple financial model in MS Excel, and developing the same model in Python which may be used as an *oracle* to test spreadsheet calculations against. With less effort than likely imagined by those unfamiliar with the language, spreadsheet developers can create a comprehensive library of Python-based tools to test spreadsheet results across numerous disciplines.

## 2. PREVOUSLY PRESCRIBED SOLUTIONS

At a single spreadsheet analysis level, Aurigemma and Panko classify spreadsheet error detection protocols across three categories: testing (a tool traditionally used in software engineering), inspection, and spreadsheet static analysis tools (SSATs).

Inspection is a particularly cost and labor-intensive process: it requires teams of employees to inspect code cell-by-cell at multiple intervals in order to prevent cognitive fatigue, and missed results. Even then, this process only yields an error detection result range of 60-80% under the most optimal of circumstances [Aurigemma and Panko, 2010].

SSATs take a different approach. As software – usually in the form of Excel add-ins – SSATs use various tools such as pattern matching and visual mapping in order topoint a user to likely problematic cells.

Across numerous tests, SSATs have shown to work in a similar degree of accuracy compared to the more labor-intensive inspection protocols, yet question remains about specifically *what* errors they perform well in detecting, and furthermore, results across tools lack in consistency. Kulesz and Ostberg describe this lack of unification in results as an issue of different tools, different configurations, different ways of showing output, but most importantly, different hidden software assumptions that provide different results for even the most-simple of patterns such as "constants in formulas" errors. Furthermore, certain SSATs run into trouble with Excel version upgrades; the authors call for tool vendors to decouple the core audit/inspection functions from the front-end (interface-level) execution environment, and they find only two tools which do so [Kulesz and Ostberg, 2013].

A third protocol prescribed by Aurigemma and Panko is the practice of *testing* where spreadsheets would be tested against calculations provided by an *oracle* model. Aurigemma and Panko note that while testing may be one of the better ways to confirm the accuracy of a model in question, the Excel environment does not provide much in the way of testing tools; as such, the testing of complex spreadsheets would require rebuilding another complex spreadsheet – an *oracle* – to test against. Again, this approach proves labor intensive. Given the combinatorial complexity of large, complex spreadsheets with many inputs and calculations performed, *oracles* developed in order to test complex spreadsheets may be prone to the same or different errors originally conceived, delivering a lack of clarity as to which spreadsheet actually provides the proper results [Aurigemma and Panko, 2010].

Shubbak and Thorne approach the spreadsheet error problem from an organizational level by developing a program which assessesrisks of spreadsheets within an organization, in an effort to provide decision-makers or auditors with the most important spreadsheets to place their focus on. A key issue, the authors note, is that of redundancy, and a lack of centralization. Due to the ease of developing spreadsheets, end users often opt to re-develop their own implementations of existing tools, rather than use an organization's previously built tool which may serve the same purpose. The lack of quick and easy clarity of someone else's spreadsheet logic, especially when sheets of a higher complexity are developed, creates redundancy within an organization's spreadsheet software repository. An alternative issue may be those issues of frequent use. The lack of a strong version control system, and/or documentation to accompany spreadsheet-based tools within the Excel framework provides a lack of clarity in where and how spreadsheets have been edited. This creates not only a lack of clarity for calculations, but a lack of clarity as to the model's input assumptions as well. [Shubbak and Thorne, 2015]

The solutions prescribed above are not without their merits, yet each method attempts to mitigate errors derived largely from a tension between the notion that "spreadsheets are software and should be developedaccording as such" and the notion that "spreadsheet-authors are not programmers and would not concern themselves with such traditional development practices" [Ayalew et al., 2000]. Implicit in almost every paper is the notion that spreadsheets are used as alternatives to programming languages due to their lack of (perceived) complexity, and ease of use.

## 3. OPEN SOURCE AS AN ALTERNATIVE TO PRESCRIBED SOLUTIONS

Recent innovations the open source programming language, Python, have made significant inroads in bridging the gap between the software development flexibility, and interactive user-friendliness. Previously, many programming languages (VBA included) have lacked a strong, user-friendly interface for interactive computing. The development of an interface called the Jupyter Notebook enables the user to run calculations in a simple, documentable fashion. While the topic of this paper will later focus on the use of the Notebook platform to develop an *oracle,* further discussion of open source technology is warranted to provide a complete discussion to the previously discussed solutions in section 2.

Shubbak and Thorne describe centralization as one possible technical solution to mitigating spreadsheet error risk, however they state that this may incur risks of its own owing in part to new environments being unsuitable to artefacts in most organizations. Additionally, they describe how when a spreadsheet is used and modified by multiple users, the risk of error increases due to the possible changes performed on the sheet [Shubbak and Thorne 2015]. Implicitly, this is a version control issue: the Excel environment lacks a strong system for tracking and managing changes. In the programming community (open or closed source), centralization and version control are the rule rather than the exception. Platforms such as Github and Bitbucket allow an enterprise to publicly or privately manage their entire codebase from a central location, with changes tracked by user, often with descriptions (depending on internal policy), at every update. These de-facto practices and environments represent a significant departure from the Excel and VBA-related environments.

The previously cited papers have concluded that SSATs work to a reasonable, yet non-comprehensive degree of accuracy. As Kulesz and Ostberg note, the applications present differences in results for even simple pattern matchings due specifically to a program's inexplicit assumptions and exceptions on how to treat various errors [Kulesz and Ostberg, 2013]. This is a problem that can be easily remedied by the open-sourcing of an audit/inspection engine to provide a base standard of unification from which third party, topic-specific solutions can be built from. Should such an open source solution be implemented, the userbase would additionally unlock access to the well-documented machine learning analytics packages which lay central to the Python language's data analysis power. In doing so, programs may better develop predictions and classifications within spreadsheets, given a proper corpus of data, whether privately by an audit company, or publicly through known spreadsheet repositories.

## 4. METHODOLOGY

### 4.1 Overview of Approach & Goals

With an end goal of developing a library of Python functions with which to test future spreadsheet models against – referenced hereafter as the *oracle*, this paper will explore a possible path of recreating spreadsheet calculations in the Python language, and then define an approach to auditing such calculations within the spreadsheet itself. While the approach below begins by building an auditing toolkit by taking a pre-existing Excel-based *oracle*(one which we assume to be correct), and replicates it in Python, this paper intends to showcase the clarity of the calculations being performed in the Python language itself. After replicating and validating the model described below, this paper will intentionally introduce an error into the spreadsheet calculations, and utilize one possible approach of identifying where this error occurs.

Both files may be downloaded from the author's website: http://www.triviumfinancialgroup.com/wp-content/uploads/2017/06/eusprig_solar.zip.

### 4.2 Example to be Used

This example will showcase time-flexible calculations of EBITDA for a solar power purchase agreement, given the following inputs:

- Monthly Avg. Solar Irradiance (kWh/m$^2$)
- Plant Capacity/Size (kWp)
- Derate Factor (%)
- Annual Plant Degradation Rate (%)
- PPA Sales Price ($/kWh)
- Operations & Maintenance Costs ($/kWh)
- Inflation Rate for PPA Escalation and Operational Expenditures (%)

### 4.3 Setting Up the Spreadsheet Model

The spreadsheet setup of this model is show below, and the following walk through will begin with the exhibit name first, followed by a description and the corresponding image.

Spreadsheet Inputs are distinguished between Model Inputs, which will apply to both the spreadsheet and the Python versioning. Spreadsheet inputs will only be required locally for the spreadsheet model. Certain inputs in Excel have been named for ease of access in Python.

*Exhibit 4.1: Model Timing with Masks and Operational Inputs*

|  | C | D | E | F | G | H | I | J | K | L | M | N |
|---|---|---|---|---|---|---|---|---|---|---|---|---|
| 2 | TRIVIUM |  |  | Key: | Model Input |  |  |  |  |  |  |
| 3 | FINANCIAL GROUP |  |  |  | Spreadsheet Input |  |  |  |  |  |  |
| 4 |  |  |  |  | Named Cell |  |  |  |  |  |  |
| 5 | Solar Ops/Python Project for EuSpRiG 2017 |  |  | Pre-Operations |  |  |  |  |  |  |  |
| 6 | **Timing/Masks** |  |  |  |  |  |  |  |  |  |  |
| 7 | Timeline |  |  |  |  |  |  |  |  |  |  |
| 8 | Model Years | 5 |  |  |  |  |  |  |  |  |  |
| 9 | Model Months |  |  | 0 | 1 | 2 | 3 | 4 | 5 | 6 | 7 | 8 |
| 10 | Operations Start Month: | 1 |  |  |  |  |  |  |  |  |  |
| 11 | Operations Mask |  |  |  | TRUE | TRUE | TRUE | TRUE | TRUE | TRUE | TRUE | TRUE |
| 12 | Operating Month |  |  | 0 | 1 | 2 | 3 | 4 | 5 | 6 | 7 | 8 |
| 13 |  |  |  |  |  |  |  |  |  |  |  |
| 14 | Generation Month w/ MOD Function: |  |  |  | 1 | 2 | 3 | 4 | 5 | 6 | 7 | 8 |
| 16 | **Solar Operations to Compute EBITDA** |  |  |  |  |  |  |  |  |  |  |
| 17 | Solar Plant/Operations Inputs |  |  |  |  |  |  |  |  |  |  |
| 18 | Plant Size (kWp) | plant_size | 500 |  |  |  |  |  |  |  |  |
| 19 | Derate Factor (x) | derate | 1 |  |  |  |  |  |  |  |  |
| 20 | PPA Price ($/kWh) | ppa_price | $0.08 |  |  |  |  |  |  |  |  |
| 21 | O&M Cost ($/kWh) | om_cost | $0.02 |  |  |  |  |  |  |  |  |
| 22 | Annual Degradation (%) | degradation | 1.0% |  |  |  |  |  |  |  |  |

## 4.4 Modeling the Net Power Generation

The solar irradiance table (G26:O37 below) shows the total irradiance (the sun's resource available to be transformed into energy by the solar plant) per period by making use of an INDEX lookup function, and is multiplied by the operating mask in row 11 of the timing section (above) for each corresponding period. While laying a month-by-calculation out in a table might be unnecessary for a monthly model, these tables quickly become useful when periods consisting of multiple months (quarters, semi-annual, etc.) are to be applied. These figures are summed in row 38, and multiplied by the plant size, and by the derate factor (the loss of power conversion from DC to AC)in order to compute Nominal Generation. Nominal Generation is then divided by a Degradation Index to compute the Net Power Generation of a project.

*Exhibit 4.2: Irradiance Table and Generation Calculations*

|  | C | D | E | F | G | H | I | J | K | L | M | N | O |
|---|---|---|---|---|---|---|---|---|---|---|---|---|---|
| 2 | TRIVIUM |  |  | Key: | Model Input |  |  |  |  |  |  |  |
| 3 | FINANCIAL GROUP |  |  |  | Spreadsheet Input |  |  |  |  |  |  |  |
| 4 |  |  |  |  | Named Cell |  |  |  |  |  |  |  |
| 5 | Solar Ops/Python Project for EuSpRiG 2017 |  |  | Pre-Operations |  |  |  |  |  |  |  |  |
| 24 | **Irradiance Scheduling** |  |  |  |  |  |  |  |  |  |  |  |
| 25 | Month | Irradiance (kWh) |  |  |  |  |  |  |  |  |  |  |
| 26 | 1 | 233.40 |  |  | 233.4 | 0 | 0 | 0 | 0 | 0 | 0 | 0 | 0 |
| 27 | 2 | 194.00 |  |  | 0 | 194 | 0 | 0 | 0 | 0 | 0 | 0 | 0 |
| 28 | 3 | 198.80 |  |  | 0 | 0 | 198.8 | 0 | 0 | 0 | 0 | 0 | 0 |
| 29 | 4 | 165.60 |  |  | 0 | 0 | 0 | 165.6 | 0 | 0 | 0 | 0 | 0 |
| 30 | 5 | 138.90 |  |  | 0 | 0 | 0 | 0 | 138.9 | 0 | 0 | 0 | 0 |
| 31 | 6 | 122.70 |  |  | 0 | 0 | 0 | 0 | 0 | 122.7 | 0 | 0 | 0 |
| 32 | 7 | 138.90 |  |  | 0 | 0 | 0 | 0 | 0 | 0 | 138.9 | 0 | 0 |
| 33 | 8 | 164.50 |  |  | 0 | 0 | 0 | 0 | 0 | 0 | 0 | 164.5 | 0 |
| 34 | 9 | 187.10 |  |  | 0 | 0 | 0 | 0 | 0 | 0 | 0 | 0 | 187.1 |
| 35 | 10 | 213.90 |  |  | 0 | 0 | 0 | 0 | 0 | 0 | 0 | 0 | 0 |
| 36 | 11 | 218.70 |  |  | 0 | 0 | 0 | 0 | 0 | 0 | 0 | 0 | 0 |
| 37 | 12 | 229.30 |  |  | 0 | 0 | 0 | 0 | 0 | 0 | 0 | 0 | 0 |
| 38 | Total: |  |  |  | 233.4 | 194 | 198.8 | 165.6 | 138.9 | 122.7 | 138.9 | 164.5 | 187.1 |
| 40 | **Generation** |  |  |  |  |  |  |  |  |  |  |  |
| 41 | Nominal Generation |  |  |  | 116700 | 97000 | 99400 | 82800 | 69450 | 61350 | 69450 | 82250 | 93550 |
| 42 | Degradation Index | 1 |  |  | 1.001 | 1.002 | 1.002 | 1.003 | 1.004 | 1.005 | 1.006 | 1.007 | 1.007 |
| 43 | Net Generation |  |  |  | 116603 | 96839 | 99153 | 82526 | 69163 | 61046 | 69048 | 81706 | 92854 |

## 4.5: Computation of Real and Nominal EBITDA

Revenue is calculated by multiplying the net power generation by a hypothetical sales price per kWh, and Operational Expenses are computed in the same manner. EBITDA = Gross Revenue – Operational Expenditures. To calculate the Nominal values, these figures are multiplied by an inflation index.

*Exhibit 4.3: Computation of Real and Nominal EBITDA*

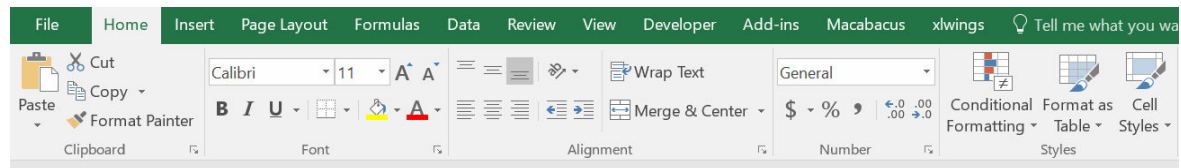

## 5. AUDIT *ORACLE* TOOLKIT CREATION & VALIDATION

**5.1 Python Package Requirements**

This presentation makes use of the Python language (version 3.0 +), and three free, readily accessible packages: xlwings, NumPy, and the Jupyter Notebook interface. For users new to Python, the easiest way to access the bundle is by going to http://continuum.io/downloads, and download the relevant open-source Anaconda distribution package. Anaconda includes all the prerequisites above, and many more.

Python, and its various packages, may be best understood by first examining a familiar exhibit: the Excel Home ribbon.

*Exihbit 5.1: Excel Home Ribbon*

Imagine Excel if all it had were basic arithmetic capacities, and perhaps simple formatting from the home tab. The capabilities provided in the Formulas, Data and Charting tabs are all still available, only they would require separate installation. This is the case with Python. By downloading only the Python language, we have installed a platform for additional functionality. Each additional package installed adds specific functionality ranging from numerical computing (NumPy package), integration with Excel (xlwings package), and an accessible user interface (IPython/Jupyter Notebooks). Additional standard packages of note for spreadsheet users, which will not be discussed in this paper, are the Pandas package for data analysis, and the Matplotlib package for charting and graphing.

**5.2 Creating the *Oracle*: Defining an Approach**

In the spreadsheet example above, a simple spreadsheet model was created to compute real and nominal EBITDA for the operations of a 500kWp solar plant. Thiswas done by organizing the model into small sections of 2-3 calculations, each building off the next. This approach will be replicated as we reconstruct the model in Python's Jupyter Notebook interface by creating a series of user defined functions with the same outputs.

## 5.3 Creating & Testing the Kit Step-by-Step

*Step 1: Open the Jupyter Notebook App and Import Required Packages*

As mentioned earlier, the two key packages (also referred to as libraries)to be utilized in this walkthrough are xlwings and NumPy. xlwingsenables a to link to any running spreadsheet instances, and access any of the workbook's artefacts. NumPyis Python's main scientific/numerical computing package, driven largely by array-wise calculations similar to those described in Swan 2016. These packages provide the basis for our calculations and modelling.

Libraries make up the open source ecosystem, and this is one of the key areas where VBA and more open source technologies begin to diverge. As the packages are imported, a "namespace" is created which enables the user to reference the library's capabilities. In the first cell below, the libraries are imported "as" abbreviations. This is standard practice and creates abbreviated namespaces within the program, allowing the user to access each object or action within the library by calling the abbreviation.

*Exhibit 5.2: Importing packages "as" abbreviated namespaces.*

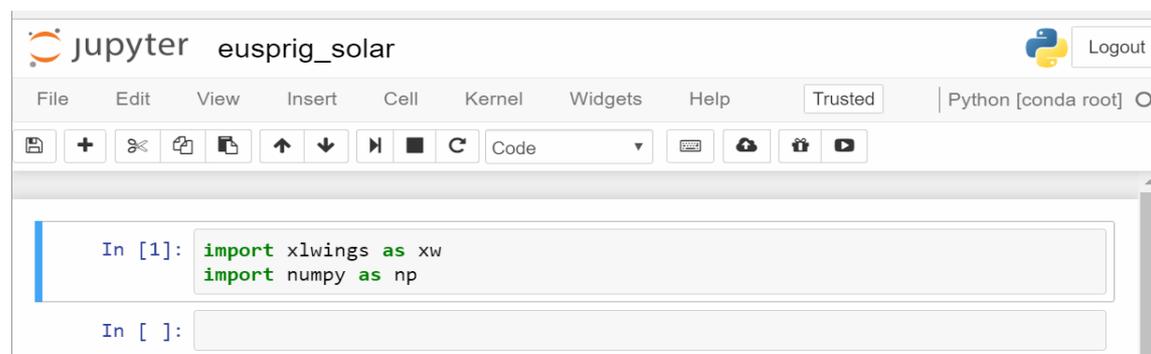

*Step 2: Create User Defined Functions*

This paper will approach modelling in Python by defining a series of functions, their respective parameters (inputs), and their returns (outputs). Math and logic come first, inputs come later. As in the Excel example, calculations are segmented into small pieces. While segmentation to this extent isn't necessary when building complete end-use models in Python, it is critical to this approach of auditing. Each of these functions can be easily reused, or combined into one, once the user is comfortable with the logic.

In addition to the function's logic, the function's use can be described with a callable "docstring". If another user wishes to use the function, but is unsure how it works, the user only needs to run a command with the function name and a question mark, which will prompt a help screen with the function's documentation. This feature applies to any documented objects, modules, or functions in Python. The ease of access of this documentation represents a significant departure from methods required by Excel or VBA.

Exhibits 5.3-5 below show the creation of a function to calculate nominal power generation with a docstring included, a call to the newly created function's documentation, and the rest of the functions required to mirror the Excel model with comments in place of full documentation for concision.

*Exhibit 5.3: Creating a function with built-in documentation.*

```python
In [2]: # nominal generation
        def nom_generation(plant_size, derate, irradiance, start_month, model_years):
            """
            This function returns the nominal AC power generation of a solar plant.

            Parameters
            ===========
            plant_size : float
                the size of the solar plant in terms of kW
            derate : float
                the conversion/loss factor between DC and AC energy
            irradiance : array
                the monthly solar irradiance in kWh/m2, in order from January to December
            start_month : integer
                the numerical month the project is expected to begin operations
            model_years : integer
                the modeled lifetime of the project in years

            return type: array
            """
            start = int(start_month - 1) # adjust for zero-indexing and integer datatype
            model_years = int(model_years)
            irradiance_1y = np.roll(irradiance, -start)
            model_irradiance = np.tile(irradiance_1y, model_years)
            return plant_size*derate*model_irradiance
```

```
In [3]: nom_generation?
```

```
Signature: nom_generation(plant_size, derate, irradiance, start_month, model_years)
Docstring:
This function returns the nominal AC power generation of a solar plant.

Parameters
===========
plant_size : float
    the size of the solar plant in terms of kW
derate : float
    the conversion/loss factor between DC and AC energy
irradiance : array
    the monthly solar irradiance in kWh/m2, in order from January to December
start_month : integer
    the numerical month the project is expected to begin operations
model_years : integer
    the modeled lifetime of the project in years
```

*Exhibit 5.4: Calling the function's documentation interactively.*

*Exhibit 5.5: Formulas documented with comments to mirror Excel model.*

```python
# discount index formula to create degradation and inflation indices
def discount_index(ops_months_array, rate):
    factors = (1+rate)**(ops_months/12)
    return factors

# net generation
def net_generation(nominal_gen, degradation):
    return nominal_gen/degradation

# real ebitda
def real_ebitda(net_gen, ppa_price, om_cost):
    revenue = net_gen*ppa_price
    expenses = net_gen*om_cost
    ebitda = revenue - expenses
    return ebitda

# nominal ebitda
def nominal_ebitda(net_gen, ppa_price, om_cost, inflation):
    revenue = net_gen * ppa_price * inflation
    expenses = net_gen * om_cost * inflation
    ebitda = revenue - expenses
    return ebitda
```

*Step 3: Define Model Inputs by Calling from Excel Workbook*

- Using xlwings, ranges can be called by named ranges, cell values, or by selecting the first value of an array, and using the expand option.
- The Excel workbook needs to be open to call values from it.
- Similar to VBA, Python is "object-oriented"; there is a defined hierarchy in objects such as *xw.Book('eusprig_solar.xlsm').sheets['Model']* where the highest level, xw is the namespace of the xlwings package previous imported, and Book is an object defined within the namespace.

*Exhibit 5.6: Linking to Excel Workbook and Importing Assumption Values*

```python
In [3]: sht = xw.Book('eusprig_solar.xlsm').sheets['Model'] #define the sheet
        irradiance = sht.range('D26').options(np.array, expand='down').value #call array
        plant_size = sht.range('plant_size').value
        derate = sht.range('derate').value
        ppa_price = sht.range('ppa_price').value
        om_cost = sht.range('om_cost').value
        deg_rate = sht.range('degradation').value
        inf_rate = sht.range('D45').value
        start_month = sht.range('D10').value
        model_years = sht.range('D8').value

        ops_months = np.arange(1, model_years*12 + 1) #add 1 due to zero-indexing
```

*Step 4: Perform Excel Model Steps in Python*

- Running the model is a matter of calling functions we created earlier.
- The Nominal EBITDA output has been printed to examine at the raw result.

- N.B. Input parameters from Step 3 (above) have been named to match the function parameters. Just as in Excel or VBA, consistency in naming conventions is of high importance.

*Exhibit 5.7: Defining Output Variables by Running Functions Created in Step 2*

```
In [5]: #generation
        nom_gen = nom_generation(plant_size, derate, irradiance, start_month, model_years)
        deg_index = discount_index(ops_months, deg_rate)
        net_gen = net_generation(nom_gen, deg_index)

        #ebtida & inflation adjustments
        ebitda_r = real_ebitda(net_gen, ppa_price, om_cost)
        inf_index = discount_index(ops_months, inf_rate)

        #could be computed easier...
        ebitda_n = nominal_ebitda(net_gen, ppa_price, om_cost, inf_index)

        print('Nominal EBITDA: ', ebitda_n)

Nominal EBITDA:  [ 7007.75117558  5829.56457816  5978.70787986  4984.34222296  4184.14120902
   3699.17788833  4191.0174196   4967.5207397   5654.62931968  6469.90202742
   6620.52255148  6947.10891089  7077.13485059  5887.28303933  6037.90300738
   5033.69214596  4225.5683497   3735.80341198  4232.51264158  5016.70411337
   5710.61574859  6533.96046333  6686.07227971  7015.89216743  7147.20549267
   5945.57297041  6097.68422528  5083.53068206  4267.40566009  3772.79156457
   4274.41870734  5066.37445112  5767.15649858  6598.65314119  6752.27101515
   7085.35644632  7217.96990349  6004.44002952  6158.05733642  5133.86266901
   4309.65720128  3810.1459365   4316.73968464  5116.5365744   5824.25705797
   6663.9863406   6819.12518362  7155.50849034  7289.43495204  6063.8899308
   6219.02820114  5184.69299246  4352.32707456  3847.87015369  4359.47968152
   5167.19535236  5881.92296944  6729.96640338  6886.64127455  7226.35510906]
```

*Step 5: Compare and Validate Results between Python Calculations and Excel*

General Approach:
- Best practice: start from the last calculation. In this case, nominal EBITDA.
- Use np.allclose(), and np.isclose() comparison functions evaluate results within a set tolerance. 1x10^5 set as default.
- Use np.where() to determine cell locations - similar to Excel's MATCH function.

Other Practical Considerations:

- The range called in Excel contains extra columns as contingencies for a delayed operating start: calculations are housed within the spreadsheet framework, whereas in Python, calculations only as long as required. To adjust for this, we need to resize the Excel array to isolate only the columns used in the calculation.
- These columns can be isolated by "slicing" larger array to appropriate size. This is done by adding brackets after an array name.
- To take the appropriate "slice", we have adjust for Python's zero-based indexing.

*Exhibit 4.8: Confirming Model Equivalency and Accuracy*

```
In [6]: ### Testing when all calculations are correct

        #call nominal EBITDA array from Excel using expand='right', or range name if used
        xl_ebitda_n = sht.range('G56').options(np.array, expand='right').value #or sht.range('ebitda_n').value

        #adjust raw Excel array for calculations as defined by model parameters
        start_index = int(start_month - 1)
        xlAdj_ebitda_n = xl_ebitda_n[start_index:len(ops_months)]

        #compare results when both models are correct
        np.allclose(xlAdj_ebitda_n, ebitda_n)

Out[6]: True
```

## 6. AUDITING CALCULATIONS

### 6.1 Finding the Error: Approach Overview

Now that the *oracle* has been created and validated, it can be used to test against erratic spreadsheets with the same goal. In this section, a common financial modelling error has been introduced to an undisclosed section of the spreadsheet model. This section will walk step-by-step through the *oracle*'s use, with help from the previously mentioned NumPy functions, allclose(), isclose() and where().

### 6.2 Using the Python *Oracle* to Find the Error: Step-by-Step

*Step 0: Replaying Our Last Calculation with New Information*

As seen below, replaying the previous validation command returns a False output. The calculations do not match up.

*Exhibit 6.1: Retesting Model Equivalency*

```
In [7]: #start at the bottom again: comparing Nominal EBITDAs using same calculations as above
        xl_ebitda_n = sht.range('ebitda_n').value #same as above, but with named range instead of expand method

        start_index = int(start_month - 1)
        xlAdj_ebitda_n = xl_ebitda_n[start_index:len(ops_months)]

        np.allclose(xlAdj_ebitda_n, ebitda_n)

Out[7]: False
```

*Step 1: Step Backwards to Find Correct Sections*

A series of print statements enables the user to "step backwards" through the major calculation points of the model to in an attempt to isolate an area where the error has been introduced. Calculations such as inflation or degradation would be skipped in this initial section as they are additive to more major breakpoints, and would be exploredonly if our initial search warranted it.

*Exhibit 6.2: Testing Model Modules with a Broad Scope*

```
In [20]: #import arrays from Excel w/ index adjustments
         xl_ebitda_r = sht.range('G51').options(np.array, expand='right').value[start_index:len(ops_months)]
         xl_net_gen = sht.range('G43').options(np.array, expand='right').value[start_index:len(ops_months)]
         xl_nom_gen = sht.range('G41').options(np.array, expand='right').value[start_index:len(ops_months)]

         #allclose() tests
         print('Real EBITDA: ', np.allclose(ebitda_r, xl_ebitda_r))
         print('Net Generation: ', np.allclose(net_gen, xl_net_gen))
         print('Nominal Generation: ', np.allclose(nom_gen, xl_nom_gen))

         Real EBITDA:  False
         Net Generation:  False
         Nominal Generation:  True
```

*Step 2: Drilling Down on Errors*

Above, the Nominal Generation calculations return True, while Net Generation remains False. This leaves two possible areas to look for drill down for errors: the Degradation Index calculation, and the Net Generation calculation.

From here, we can run two tests. As the degradation index has not yet been compared, we will call the np.allclose() function once more. As Jupyter's interface allows for multiple commands at a time, Net Generation can be examined further in the same block further by testing for similarity between each element in the array (as seen below).

*Exhibit 6.3: Testing with a Higher Level of Detail*

```
In [21]: xl_deg_index = sht.range('G42').options(np.array, expand='right').value[start_index:len(ops_months)]

         # allclose() for degradation index
         print('Degradation Index: ', np.allclose(deg_index, xl_deg_index))

         # isclose() to find problem area of net generation
         print('Net Generation: ', np.isclose(net_gen, xl_net_gen))

         Degradation Index:  True
         Net Generation:  [ True False False False False False False False False False False False
          False False False False False False False False False False False False
          False False False False False False False False False False False False
          False False False False False False False False False False False False
          False False False False False False False False False False False False]
```

As results above are examined, the outputs show that the Degradation Index is not the source of error. The Net Generation np.isclose() output shows that while the first "cell" in the array is correct, the rest are not. By drilling down on initial tests, the source of the error has been identified.

This error is immediately guessable: the first cell in an array was edited, but not unlikely pasted across.

*Step 3: Identifying Problematic Excel Cells*

Above, results showed that the first cell of the Net Generation array is correct, and likely not pasted across. This is easy to see visually because it occurs at the beginning of the array. But what if it was somewhere in the middle of the array? By calling the Excel range object of the array, without appending it with a ".value" method, we can use the np.where function and some indexing (as seen below in Exhibit 6.4) to return the correct and erratic cells of the array.

*Exhibit 6.4: Identifying the Error Cell Locations*

```
In [36]: xl_nom_gen_rng = sht.range('G41').expand('right')

         #find index locations
         net_gen_isclose = np.isclose(net_gen, xl_net_gen)
         error_indices = np.where(net_gen_isclose == False)
         correct_indices = np.where(net_gen_isclose == True)
         print('Correct Indices: ', correct_indices)
         print('Error Indices: ', error_indices)

         #find the appropriate indices beginning and end
         correct_ix = correct_indices
         error_ix_max = max((np.amax(error_indices)), len(ops_months))
         error_ix_min = np.amin(error_indices)

         correct_range = xl_nom_gen_rng[:, correct_ix_min:correct_ix_max].address
         error_range = xl_nom_gen_rng[:, error_ix_min:error_ix_max].address

         print('Correct Cell/Range: ', correct_range)
         print('Error Cell/Range: ', error_range)

         Correct Indices:  (array([0], dtype=int64),)
         Error Indices:  (array([ 1,  2,  3,  4,  5,  6,  7,  8,  9, 10, 11, 12, 13, 14, 15, 16, 17,
                18, 19, 20, 21, 22, 23, 24, 25, 26, 27, 28, 29, 30, 31, 32, 33, 34,
                35, 36, 37, 38, 39, 40, 41, 42, 43, 44, 45, 46, 47, 48, 49, 50, 51,
                52, 53, 54, 55, 56, 57, 58, 59], dtype=int64),)
         Correct Cell/Range:   $G$41{1x0}
         Error Cell/Range:   $H$41:$BN$41
```

## 6. CONCLUSIONS

This paper has sought to provide the foundations for developing an alternative to building test *oracles* in MS Excel by laying the groundwork in the Python language for a predominantly spreadsheet-oriented audience. Embedded in this paper is the notion that while Excel might be the current de facto tool for financial modelling, less risky alternatives exist which – when used appropriately – can replicate and build on many capabilities traditionally found in Excel.

The numerous steps shown in this paper could be safely, and concisely shortened. While the current lack of adoption among the financial community is a hindrance, the Python language presents itself as a strong alternative to the risks associated with spreadsheet model risk due to recent innovations in usability such as the Jupyter Notebook interface, and the language's highly legible syntax. Calculations are clear, wordiness is low, and the amount of programming knowledge required to reconstruct Excel-oriented analysis tools is accessible to the average spreadsheet professional.